\documentclass[article]{IEEEtran}
\usepackage{graphicx}    
\usepackage{subfigure}
\usepackage{cite}
\usepackage[T1]{fontenc}
\usepackage{bm}
\usepackage[cmex10]{amsmath}
\usepackage{amssymb}
\usepackage[numbers]{natbib}
\usepackage[version=3]{mhchem}
\setcitestyle{aysep={},square,citesep={,}}
 
\ifCLASSINFOpdf
\else
\fi
\numberwithin{equation}{section}
\def\longrightharpoonup{\relbar\joinrel\rightharpoonup}
\def\longleftharpoondown{\leftharpoondown\joinrel\relbar}

\def\longrightharrowpoons{
\mathop{
    \vcenter{
      \hbox{
      \ooalign{
        \raise2pt\hbox{$\longrightarrow\joinrel$}\crcr
	  			}
      }
    }
  }
}
\def\longrightleftharpoons{
  \mathop{
    \vcenter{
      \hbox{
      \ooalign{
        \raise1pt\hbox{$\longrightharpoonup\joinrel$}\crcr
	  \lower1pt\hbox{$\longleftharpoondown\joinrel$}
	  }
      }
    }
  }
}

\newcommand{\rates}[2]{\displaystyle
  \mathrel{\longrightleftharpoons^{#1\mathstrut}_{#2}}}
\newcommand{\rate}[1]{\displaystyle
  \mathrel{\longrightharrowpoons^{#1\mathstrut}}}
	
\begin{document}
	\title{A Low Dimensional Approximation For Competence In Bacillus Subtilis}
	\author{\IEEEauthorblockN{An Nguyen, Adam Pr\"{u}gel-Bennett, Srinandan Dasmahapatra} \\
	\IEEEauthorblockA{Faculty of Physical Sciences and Engineering\\
    University of Southampton\\
    Southampton - United Kingdom\\
    Email: ancntt2002@gmail.com}
    }	

	\maketitle
	\IEEEdisplaynontitleabstractindextext {
	\begin{abstract}	
	The behaviour of a high dimensional stochastic system described by a Chemical Master Equation (CME) depends on many parameters, rendering explicit simulation an inefficient method for exploring the properties of such models. Capturing their behaviour by low-dimensional models makes analysis of system behaviour tractable. In this paper, we present low dimensional models for the noise-induced excitable dynamics in \emph{Bacillus subtilis}, whereby a key protein ComK, which drives a complex chain of reactions leading to bacterial competence, gets expressed rapidly in large quantities (competent state) before subsiding to low levels of expression (vegetative state). These rapid reactions suggest the application of an adiabatic approximation of the dynamics of the regulatory model that, however, lead to competence durations that are incorrect by a factor of 2. We apply a modified version of an iterative functional procedure that faithfully approximates the time-course of the trajectories in terms of a 2-dimensional model involving proteins ComK and ComS. Furthermore, in order to describe the bimodal bivariate marginal probability distribution obtained from the Gillespie simulations of the CME, we introduce a tunable multiplicative noise term in a 2-dimensional Langevin model whose stationary state is described by the time-independent solution of the corresponding Fokker-Planck equation.  
	\end{abstract}
	
}
\section{Introduction}
\label{sec:intro}
\begin{figure}[h]
\centering
		\includegraphics[width=0.5\textwidth]{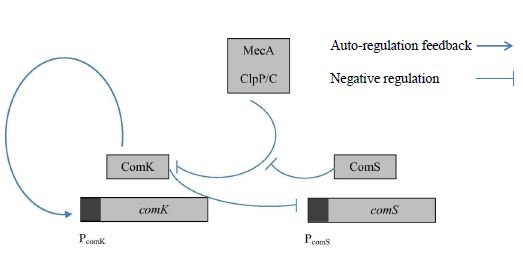}
	\caption{Competence circuit architecture. The competence circuit includes the following components: two genes $comK$ and $comS$ corresponding to two proteins ComK and ComS, respectively; and promoters $P_{comK}$ and $P_{comS}$. In this figure, ComK actives the expression of its own gene (auto-regulation feedback) and inhibits expression of ComS (negative regulation), that in turn interferes with degradation of ComK. The complex of MecA, ClpP/C also actively degrades ComK.}
	\label{fig:comcirarch}
\end{figure}

The speed and quantity of production of functional proteins to respond to the needs of an organism is regulated by transcription factor proteins that modulate the efficiency of mRNA transcript generation from which the proteins are translated, proteases that regulate their degradation, and a host of other molecular agents. Using fluorescent markers in cell-sorting \citep{Shimomura62,Herzenberg02,Lippincott03} and time-lapse recording technologies \cite{Hinchcliffe05}, dynamic traces of the cellular behaviour have become accessible, making it possible to develop a quantitative understanding of the dynamics of gene regulation in single cells. In starved conditions, some \emph{Bacillus subtilis} cells undergo a transformation called competence, ingesting DNA from its environment. Single cell fluorescent images reveal bimodal cell populations of ComK, a protein that is used to track the competent state. Bimodal probability distributions can be generated by an underlying stochastic dynamical system whose deterministic state is bistable. Maamar and Dubnau~\cite{Maa05} demonstrate that the positive feedback provided by ComK proteins activating its own transcription can generate switching behaviour, whose noise-induced activation yields bimodal distributions. An alternative model has been proposed by S\"{u}el et al.~\citep{Suel06, Suel07} which includes a slower, negative influence on ComK levels via the expression of the \emph{comS} gene, shown in Fig.~\ref{fig:comcirarch}. This leads to an excitable system~\cite{Lindner04}, whose high expression (competent) state is not stable, but undergoes slow decay back to the low-expression (vegetative) state. The long-lived state thus explains the bimodality and removes the need to invoke a mechanism for the loss of competence that the bistable model requires. This is the model that has been well studied and has received considerable attention.     
\par
The model we study ~\cite{Suel06} represented as a network graph in Fig.~\ref{fig:comcirarch}, translates into a mathematical description involving multiple reaction and species types in a continuous time Markov process, the Chemical Master Equation (CME)\cite{Gill07}. The mean values of the random variables are described by a system of ordinary differential equations. While the full stochastic description (high-dimensional) is faithful to the kinds of processes believed to occur in a cell, insights into the dynamical behaviour of these systems are reliably extracted primarily from their reduced models (low-dimensional). The excitable behaviour of the competence circuit~\cite{Suel06}, and its extended design synthesized in~\cite{Cag09}, are instances of the application of insights gained from low-dimensional non-linear dynamical systems. In ~\citep{Suel06, Suel07}, a two-dimensional reduction of the detailed model formed the basis of partitioning the phase space of the model into distinct dynamical regimes, to which the competent and vegetative states were allocated places. By altering the network properties away from its wild-type state, the authors provided evidence of the validity of their modelling framework when the engineered strain of \emph{B. subtilis} followed the model prediction, for example, entering an oscillatory phase\cite{Suel07}. However, the two-dimensional model proposed in~\cite{Suel07} fails to reproduce two salient features of the stochastic model described by the CME. The duration of the competence state in the two-dimensional model is about four hours, while those in the stochastic simulations which match the observations from time-lapse data last around 10 hours. Further, the location of the second (non-steady state) mode of the probability distribution of the ComK and ComS proteins is at a different location in the reduced model compared to the stochastic simulation.        
\par
The contribution of this paper is multifaceted but subtle. We use the method described in \citep{Fraser88,Roussel97,Roussel01} to obtain a more accurate low dimensional reduction for the existing model. This introduces a better approximation for slow-varying variables involving the binding of ComS to MecA, which corrects for non-adiabaticity in ComS dynamics. Although the underlying model parameters have enough uncertainty that this change in the competence behaviour could also be obtained by changing the values of the model parameters, it is nevertheless important in mathematical modelling to obtain a good match in the observed behaviour of the chemical master equation and its low dimensional model. A second contribution of the paper is to build a low dimensional stochastic model of the system. This is essential to capture the competence behaviour which is stochastically driven. We will show that the usual assumption of a Poissonian noise (where the variance is proportional to the number of reactions) provides a very poor description of the actual system. In contrast, we show that using a model in which the variance in the noise is proportional to the square of the molecular number can give a more accurate low dimensional approximation. In particular, by fitting the level of the noise we show that we can obtain a Fokker-Planck description which closely matches that observed in the full CME.      
\par
In order to build a two-dimensional reduced model, we identify one fast species of the two bound complexes of the protease MecA to ComK and ComS, the principal variables in the model. This is in contrast to \citep{Suel06, Suel07}, where both these complexes are assumed incorrectly, to have fast-decaying transients. However, as the ComK is not governed by exclusively fast reaction, we cannot use the adiabatic approximation \cite{Schu07}. Thus, we use an iterative scheme proposed in \citep{Fraser88,Roussel97,Roussel01} to describe the low-dimensional slow manifold as a function of the ComK and ComS variables. The reason for this is because the dynamics of the whole system breaks up into different time scales. Thus, although the individual species do not have very different time scales, there still exists one slow time scale describing the trajectory of the system after the transition to competence. This trajectory is uniquely characterized by the values of ComK and ComS. By doing this, we find that this model captures the dynamics of the decay of the competent state accurately. In order to account for the noise-driven transition to the competent state, the standard assumption is that for each reaction that contributes to the rate of change of a dynamical variable, the equality of the variance in the number of reactions with its mean, that is true for the Poisson process updates, can be used to replace the Poisson processes by independent Gaussian noise contributions with the same mean and variance. This assumption underlies the Langevin approximation~\citep{Gill02, Gill07, Cazz06} to the CME, and we find that it fails to reproduce the distributions obtained from the Gillespie simulation. Therefore, we introduce a noise term in the Langevin description  that reflects the ratio of variances of the ComK and ComS distributions at the steady state. We then tune the magnitude of the noise so that the stationary distribution, as computed from the solution of the time-independent Fokker-Planck equation, gives rise to a bimodal distribution of the ComK-ComS variables that is qualitatively similar to the marginal distribution computed from the Gillespie simulation.
\section{A MODEL FOR EXCITABLE DYNAMICS FOR EXPRESSION OF \textsc{ComK}}
\label{sec:discretestomodel}
In this section, we describe the chemical reactions provided in \citep{Suel06, Suel07} that give rise to the behaviour sought to match the observation of stochastically triggered transient competence events that are identified with high expression levels of ComK. This set of chemical reactions describe a continuous time Markov process, called a Chemical Master Equation (CME)~\cite{Gill07} which updates the probabilities of the number of molecules of each species in the system. The averages of these molecular numbers over multiple realizations is described by a set of ordinary differential equations (ODEs), called reaction-rate equations (RRE)~\cite{Gill07} whose phase portrait best illustrates the excitable dynamics~\citep{Suel06, Suel07}.
\subsection{RELATED WORK: The reactions in the discrete model (CME)}
The competence circuit includes the following components: The two principal proteins modelled are ComK and ComS and their corresponding promoters are $P_{comK}$ and $P_{comS}$. ComK inhibits expression of ComS (rate $g$) that in turn interferes with degradation of ComK by competitively binding to MecA (rate $k_{13}$); the complex of MecA also actively degrades ComK ($k_{11}$). The transcription of mRNAs occur in a basal (rates $k_1$,$k_4$) and regulated fashion (rates $f$, $g$), the stochastic reactions are described as follows (for simplicity, we denote the species concentrations $[mRNA_{comK}]$, $[mRNA_{comS}]$, $[ComK]$, $[ComS]$, $[MecA]$, $[MecA_K]$, $[MecA_S]$, as $R_K$, $R_S$, $K$, $S$, $A$, $M_K$, $M_S$):
\begin{equation}
\begin{split}
{\rm P_{comK}^{const}} &\xrightarrow{k_1} {\rm P_{comK}^{const}} + { \rm R_K}  \\     
	{\rm P_{comK}} &\xrightarrow{f\left( K,k_2,k_k\right)} {\rm P_{comK}} + {\rm R_K} \\
	{\rm R_K} &\xrightarrow{k_3} {\rm R_K} + {\rm K} \\
	{\rm P_{comS}^{const}} &\xrightarrow{k_4} {\rm P_{comS}^{const}} + { \rm R_S}  \\
	{\rm P_{comS}} &\xrightarrow{g\left( K,k_5,k_s\right)} {\rm P_{comS}} + {\rm R_S} \\
	{\rm R_S} &\xrightarrow{k_6} {\rm R_S} + {\rm S} \\
	{\rm R_K} &\xrightarrow{k_7} \emptyset \hspace{10 mm} {\rm K} \xrightarrow{k_8} \emptyset    \\
	{\rm R_S} &\xrightarrow{k_9} \emptyset \hspace{10 mm} {\rm S} \xrightarrow{k_{10}} \emptyset \\
	{\rm A} + &{ \rm K} \rates{k_{\rm 11}/\Omega}{k_{\rm -11}} {\rm M_K} \hspace{10 mm} {\rm A} + { \rm S} \rates{k_{\rm 13}/\Omega}{k_{\rm -13}} {\rm M_S}    \\  
	{\rm M_K} &\rate{k_{\rm 12}} {\rm A}  \hspace{10 mm} {\rm M_S} \rate{k_{\rm 14}} {\rm A} \\
\end{split}
\end{equation}
$P_{comK}^{const}$ and $ P_{comK}$ are constitutive and regulated promoters of ComK, respectively. $R_K$ and $R_S$ are mRNA molecules from which proteins ComK and ComS are translated, respectively. The symbols above the arrows denote the probabilities of reactions in unit time. The gene regulation functions $f$, $g$ are of the Hill type:
\begin{align*}
f\left( K,k_2,k_k\right) = \frac{k_2K^2}{{k_k}^2 + K^2} \\
g\left( K,k_5,k_s\right) = \frac{k_5}{1+\left(\frac{K}{k_s} \right)^5}
\end{align*}
The rates in the above equations are defined via the average rates of change as they appear in the RRE, and are functions of concentrations --- the number of molecules per unit volume. For bimolecular reactions, the probability of one molecule to find another is inversely proportional to the cell volume and is tracked by the parameter $\Omega$. By using the same chemical convention used in~\cite{Suel07}, then $\Omega \approx 1 nM$ (nano-molar); therefore, we can treat the concentrations of species in the same way as their molecular number by measuring concentrations in units of $nM$. 
\begin{table*}[ht]
\renewcommand{\arraystretch}{1.3}
  \caption{
	Parameters of the discrete model (source from \cite{Suel07})
	}
 \label{tbl:parameters}
\begin{center}
{
   \begin{tabular}{|c|c||c|c||c|c|} \hline
   $k_1$ & $0.00021875 s^{-1}$ & $k_7$ & $0.005 s^{-1}$  & $k_{12}$ & $0.05 s^{-1}$ \cr \hline
   $k_2$ & $0.1875 s^{-1}$ &  $k_8$  &   $10^{-4} s^{-1}$    &$k_{13}$ &   $4.5 \times 10^{-6} {nM}^{-1}s^{-1}$ \cr \hline
   $k_3$ & $0.2 s^{-1}$  &  $k_9$  &   $0.005 s^{-1}$     &   $k_{-13}$  & $5 \times 10^{-5}s^{-1}$ \cr \hline
   $k_4$ & $0 s^{-1}$  &  $k_{10}$  &   $10^{-4} s^{-1}$   &  $k_{14}$  & $4 \times 10^{-5}s^{-1}$ \cr \hline
   $k_5$ & $0.0015 s^{-1}$  & $k_{11}$ & $2.02 \times 10^{-6} {nM}^{-1}s^{-1}$  &   $k_k$  & $5000  nM$ \cr \hline
   $k_6$ & $0.2 s^{-1}$  &  $k_{-11}$  & $ 5 \times 10^{-4} s^{-1}$ &  $k_s$  & $833  nM$ \cr \hline
   \end{tabular}
 }

\end{center}
\end{table*}
\begin{table*}[ht]
\renewcommand{\arraystretch}{1.3}
\caption{Initial conditions}
  \label{tbl:initcondition}
\begin{center}
{
   \begin{tabular}{|c|c||c|c||c|c|} \hline
   $[P_{comK}^{const}]$ & $1 nM$ & $R_K$ & $1000 nM$  & $M_S$ & $100 nM$ \cr \hline
   $[P_{comS}^{const}]$ & $1 nM$ &  $R_S$  &   $1000 nM$    &K &   $1000 nM$ \cr \hline
   $[P_{comK}]$ & $1 nM$  &  A  &   $300 nM$     &   S  & $100 nM$ \cr \hline
   $[P_{comS}]$& $1 nM$  &  $M_K $ &   $100 nM$   && $$ \cr \hline
   \end{tabular}
 }

\end{center}
\end{table*}
\par
The deterministic description of the system are described by the following differential equations (the discrete model parameters are given in Table \ref{tbl:parameters}):
\begin{equation}
\begin{split}
\label{eq:7d_deterministic_model}
\frac{dR_K}{dt} &= k_1 + \frac{k_2K^2}{{k_k}^2 + K^2} - k_7R_K  \\
\frac{dR_S}{dt} &= k_4 + \frac{k_5}{1+(K/k_s)^5} - k_9R_S \\
\frac{dA}{dt} &= -k_{11}KA + (k_{-11}+ k_{12})M_K - k_{13}SA  \\
& \quad + (k_{-13}+k_{14})M_S  \\
\frac{dM_K}{dt} &= k_{11}KA - (k_{-11}+k_{12})M_K  \\
	\frac{dM_S}{dt} &= k_{13}SA - (k_{-13}+k_{14})M_S  \\
\frac{dK}{dt} &= -k_{11}KA + k_{-11}M_K + k_3R_K - k_8K  \\
\frac{dS}{dt} &= -k_{13}SA + k_{-13}M_S + k_6R_S - k_{10}S. \\
\end{split}
\end{equation}
\begin{figure}[h]
\centering
		\includegraphics[width=0.5\textwidth]{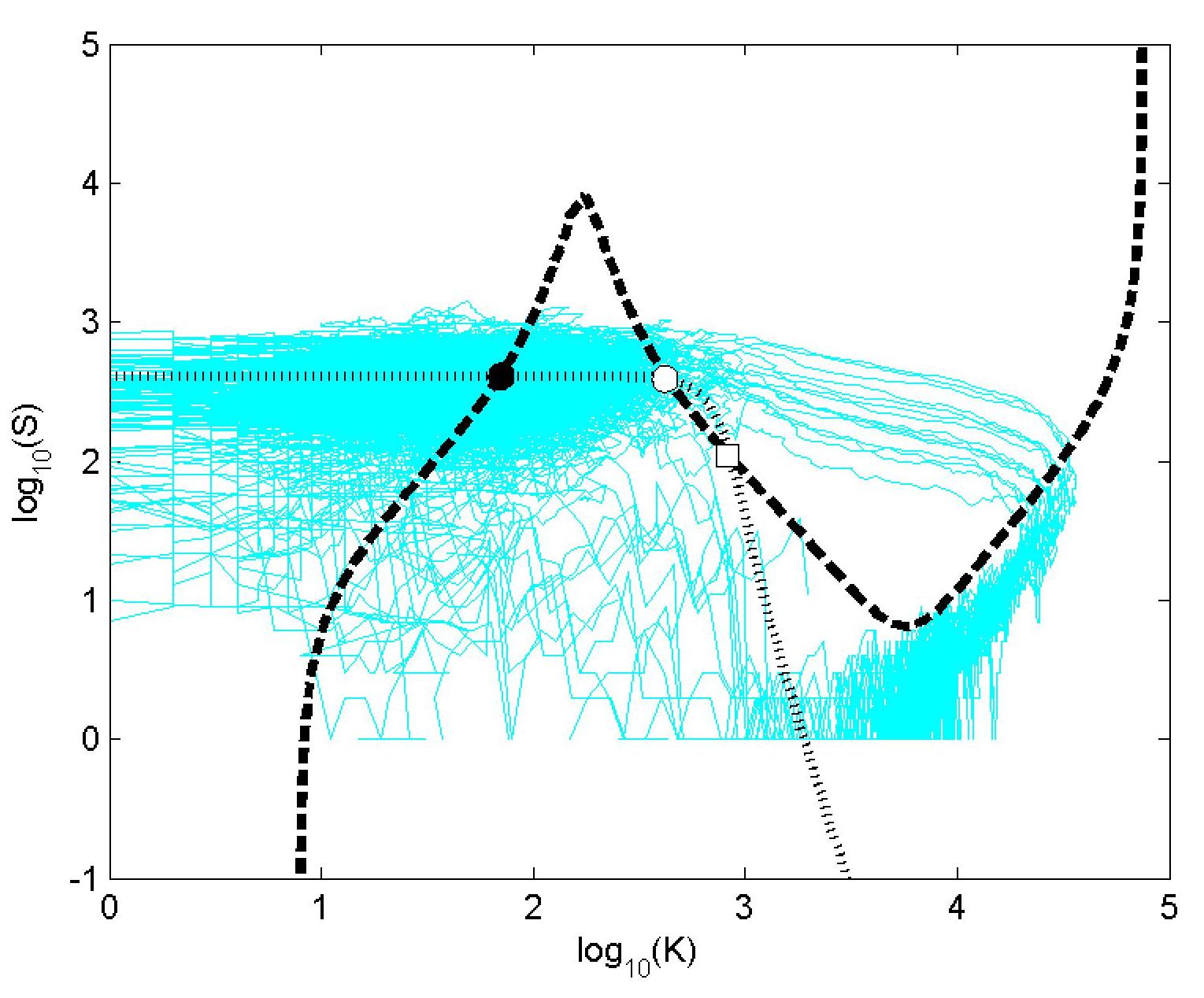}
	\caption{Trajectories generated by the Gillespie algorithm are shown in thin lines. Also shown are the nullclines of ComK (dash line), ComS (dotted line), and fixed points which are the intersections of nullclines. Stable fixed point is denoted by full circle, saddle point by empty circle, and other unstable fixed point by empty square.}
	\label{fig:trajdismod}
\end{figure}
\par 
The trajectories generated by the CME can be simulated by the Gillespie algorithm~\cite{Gill07}. Fig.~\ref{fig:trajdismod} shows such trajectories (after the initial transient) in thin lines plotted on the log-scale phase plane, for the model parameters and initial conditions in Tables \ref{tbl:parameters}, \ref{tbl:initcondition}. In addition, we plot the ComK and ComS nullclines defined by $dK/dt = 0$ and $dS/dt = 0$ with dash and dotted lines, respectively. Their intersections are the fixed points of the system dynamics. There are three fixed points, one of which is stable (full black circle), corresponding to low levels of $K$; the middle one is a saddle fixed point (empty circle). The intersection of nullclines at the highest value of $K$ corresponds to an unstable fixed point (empty square) and explains why the trajectories do not appear in the centre of the figure. The dense area around the stable fixed point is called the vegetative state and the trajectories that extrude into large ComK expression levels correspond to the competent state of the organism. 
\par
There are two critical properties of the system dynamics we are interested in: the first one is the competence duration which is the time spent by a trajectory in the state where ComK has high level of expression, which we take to be ComK numbers above $10^4$ (this threshold value is the same as that was used in ~\citep{Suel06}; therefore, our results are directly comparable with their findings); the second is the stationary probability distribution which describes the probability of the system being at a particular state. In order to understand the system dynamics, a phase plane analysis was carried out in~\citep{Suel06, Suel07} as is customary~\cite{Strogatzbook}. However, the reduced 2D model, based on an adiabatic approximation where all variables except K and S proteins are eliminated, has a time course in the competence regime that is far from that produced by the full model in (\ref{eq:7d_deterministic_model}). This is shown in Fig.~\ref{fig:comp_duration_2DDeapprsys}, where the competence duration is a factor of two smaller in this adiabatic model than in the full 7D model. In the next section, we present a method of approximating our full system by a two-dimensional one while retaining the system's competence duration. In this method, we first eliminate variables which are seen as fast variables in order to reduce the 7D system down to a 3D system. Next, we continue to reduce the 3D system to a 2D system by using an iterative procedure to capture the slow manifold of the system.
\subsection{THREE-DIMENSIONAL APPROXIMATE SYSTEM}
In order to do the model reduction, we first realize that the dynamics of mRNAs are much faster than that in the protein; therefore, we can eliminate the mRNAs by setting them to their steady state values ($dR_{K,S}/dt=0$). We restrict our attention, in this section, to the competent regime. Furthermore, since the decay rate $k_{12}$ of $M_K$ is 500 times and 10 times faster than the proteins and mRNAs respectively (notice that even though the binding/unbinding rates of MecA to ComK ($k_{11}/k_{-11}$) and ComS ($k_{13}/k_{-13}$) are similar, the decay rate of the complex $M_K$ is much faster than that of $M_S$, thus the dynamics of binding of MecA to ComK is treated different from its binding to ComS). We therefore can approximate the dynamics of $M_K$ by slaving it to the other variables. That is, we replace it by the steady state value ${M_K}^*$ obtained from solving $dM_K/dt \approx 0$, which gives ${M_K}^* = AK/\Gamma_k$, with $\Gamma_k = \frac{k_{-11}+k_{12}}{k_{11}}$. Using the conservation equation $M_T = A + M_K + M_S$, we obtain (for $Q:= A + M_K$)
\begin{equation*}
M_K = \frac{K}{\Gamma_k+K}Q
\end{equation*}
(There is some freedom in choosing which variables to eliminate and which to keep. Although this leads to the same 3D approximation, in making a further reduction to a 2D, we find that using $Q$  rather than $A$ leads to a series of approximation which converges faster to the behaviour of the 3D system. Intuitively, variable $Q$ represents an effective concentration of MecA for the ComK dynamics, the fraction that is not taken up by ComS).
\par
After making these approximations, the 3D system can be described by the following differential equations:
\begin{equation}	
\begin{split}
\label{eq:threedimeq}
	\frac{dK}{dt} &= \frac{k_3}{k_7} \left(k_1 + \frac{k_2K^2}{k_k^2 + K^2} \right) - \frac{k_{12}KQ}{\Gamma_k+K} - k_8K\\
	\frac{dS}{dt} &= \frac{k_5k_6/k_9}{1+(K/k_s)^5} - k_{10}S - k_{13}\Gamma_k\frac{SQ}{\Gamma_k+K} \\
	& \quad + k_{-13}(M_T-Q)\\
	\frac{dQ}{dt} &= (k_{14}+k_{-13})(M_T-Q) - k_{13}\Gamma_k\frac{SQ}{\Gamma_k+K}
\end{split}
\end{equation}
\begin{figure}
\centering
		\includegraphics[width=0.5\textwidth]{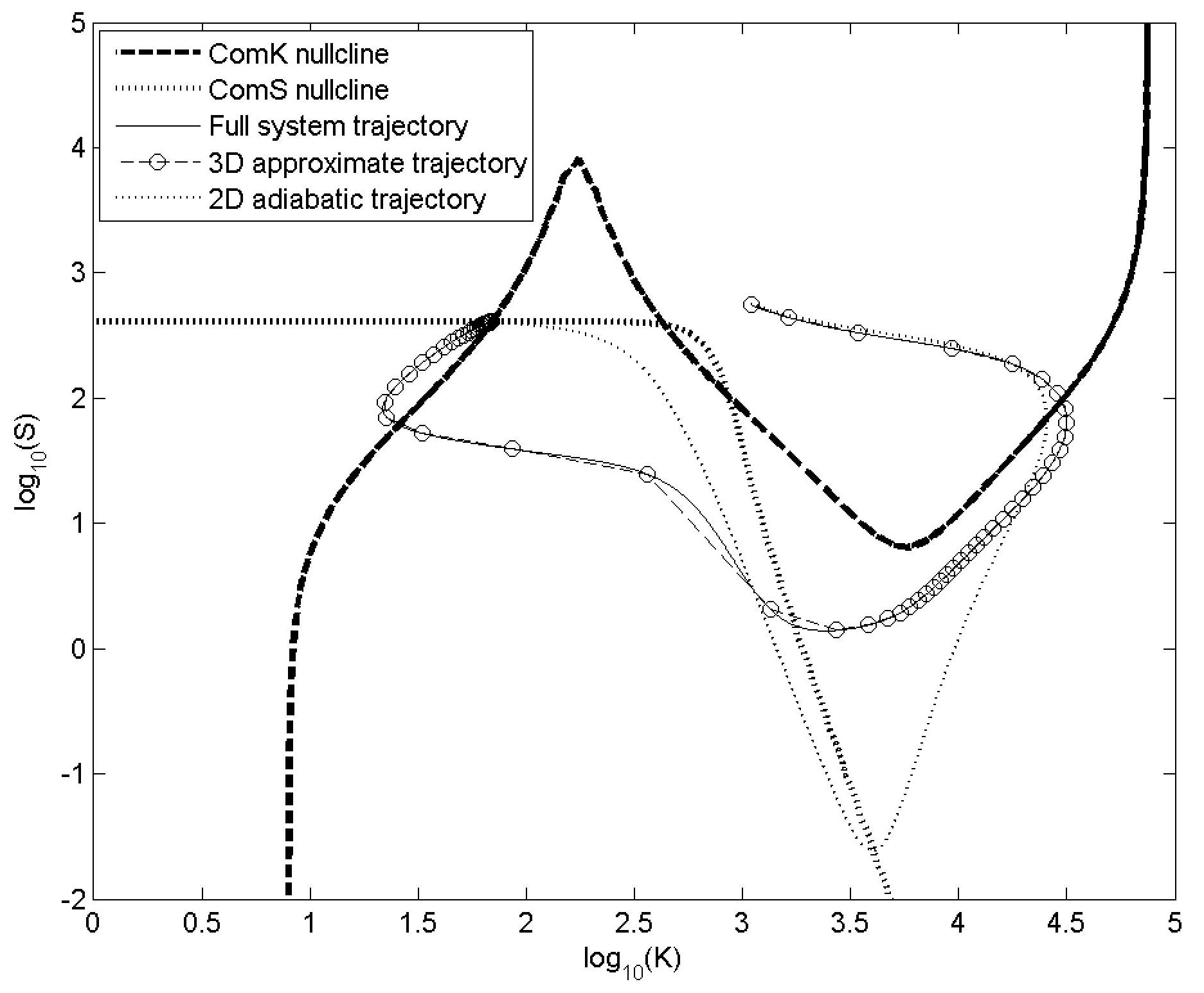}
	\caption{Trajectories in the approximate and full systems together with the nullclines of the 2D adiabatic approximate model. We compare the trajectories generated by the 3D approximate and the full models given by (\ref{eq:threedimeq}) and (\ref{eq:7d_deterministic_model}), respectively. The numerical initial condition for the integration is $K = 1099$, $S = 564$, $A = 16$, $M_K = 3$, $M_S = 481$, $R_K = 1$, $R_S = 0$. The trajectory generated by the 2D adiabatic approximate model is also shown for comparison, where it does not match that in the full system. The trajectories of the 3D approximate model are close to the ComK-nullcline, this implies that this model has captured the slow manifold in the competence regime.}
	\label{fig:syscomp}
\end{figure}   
\begin{figure*}[!t]
\centering
\mbox{
\subfigure[][]{
		\includegraphics[width=0.5\textwidth]{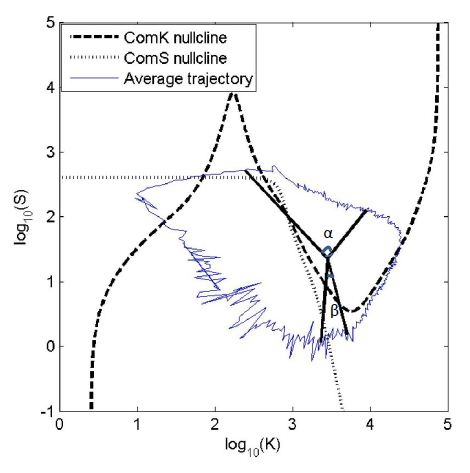}
		\label{fig:average_trajectory}
} \quad
\subfigure[][]{
	\includegraphics[width=0.5\textwidth]{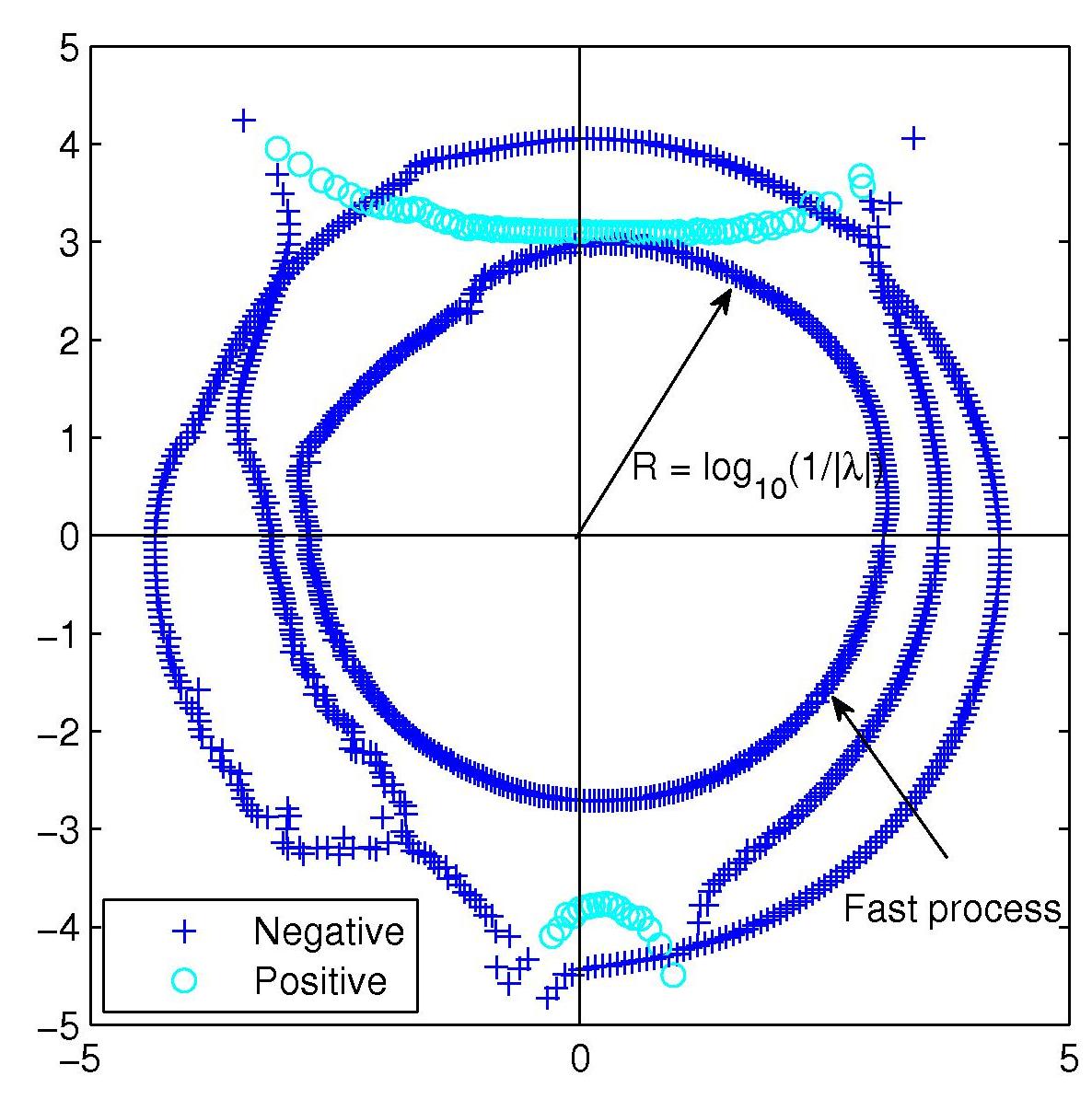}
	\label{fig:polar_axis2} 
}}
\caption{The spectra of eigenvalues on a 10-based logarithm scale. The eigenvalues are computed along the average trajectory \subref{fig:average_trajectory}, the position of the eigenvalue point is defined by the angle formed by the data point at which the eigenvalue is evaluated and the vertical axis of the polar coordinates, and the distance from that point to the origin. This distance is computed by taking the logarithm of the inverse absolute eigenvalue \subref{fig:polar_axis2}.}
\label{fig:eigenphase}
\end{figure*}
From now on, we will take the initial condition to be $K = 1099$, $S = 564$, $A = 16$, $M_K = 3$, $M_S = 481$, $R_K = 1$, $R_S = 0$ (these values come from a CME simulation). Fig.~\ref{fig:syscomp} shows that the trajectories from this 3D model (\ref{eq:threedimeq}) closely follow those from the 7D model (\ref{eq:7d_deterministic_model}). We now use the 3D model to find an approximation using a 2D model. The adiabatic approximation used in~\citep{Suel06, Suel07} is one such 2D model which gave the same fixed points as the 7D system by construction, but which gave rise to competent behaviour that was smaller by a factor of 2. This discrepancy is also shown in Fig.~\ref{fig:syscomp}. This implies that the approximation $dQ/dt\approx 0$ used there assumes that the species $Q$ changes faster than it really does in the 7D case. Moreover,  the complexes $M_K$ and $M_S$ are treated in identical ways in ~\citep{Suel06, Suel07} even though their decay rates of those species are significantly different. Also in Fig.~\ref{fig:syscomp}, the 7D model trajectories are close to the ComK-nullcline of the 2D adiabatic model, yet the 2D adiabatic model trajectories are not. The 3D approximate model shows that these trajectories and the ComK-nullcline are close to each other. This means the 3D approximate model has captured the slow manifold in the competence regime. In the following section, we first identify the existence of fast modes in the system which justifies the search for a low-dimensional approximation, and then introduce a procedure to find one.
\subsection{A TWO-DIMENSIONAL APPROXIMATE SYSTEM}
As can be seen in (\ref{eq:threedimeq}), there are no obvious fast variables; not the $Q$-variable --- the factor $(k_{14}+k_{-13})/k_{13}$ in the $Q$-evolution equation appears also in the $S$ evolution equation, but its magnitude is much smaller than that of $k_{12}$ that appears in the $K$ equation --- and not any of the others. This suggests the possibility of fast modes that decay rapidly leaving the long-time dynamics dependent on a reduced subset, which we choose to be $K$ and $S$. In order to verify this, we plot the magnitudes of the eigenvalues of the Jacobian along the average trajectory to see if they are widely separated. These are plotted in polar coordinates in the $K$-$S$ plane in Fig.~\ref{fig:eigenphase} where we have chosen an arbitrary point in the competence region as centre. For a dynamical system $d\mathbf{X}(t)/dt = f(\mathbf{X}(t))$ with variables $\mathbf{X}(t)=(K(t), S(t), Q(t))$, the linearisation of the evolution equations around each point $\mathbf{X}(t)$: $d(\delta \mathbf{X}(t))/dt = f(\mathbf{X} + \delta \mathbf{X}) - f(\mathbf{X}) = \mathbf{J} \delta \mathbf{X}(t)$ defines the Jacobian matrix $\mathbf{J}$ with matrix elements ${J}_{i,j} = \frac{\partial f_i(\mathbf{X})}{\partial X_j}$. The Jacobian matrix for the approximate system (\ref{eq:threedimeq}) is defined as,
\[
\mathbf{ J} = \bordermatrix {\text{} & (K) & (S) & (Q)\cr
(K) & \frac{\partial {f_1}}{\partial K} & \frac{\partial {f}_1}{\partial S} & \frac{\partial {f}_1}{\partial Q}\cr 
(S) &\frac{\partial {f_2}}{\partial K} & \frac{\partial {f}_2}{\partial S} & \frac{\partial {f}_2}{\partial Q}\cr 
(Q) &\frac{\partial {f_3}}{\partial K} & \frac{\partial {f}_3}{\partial S} & \frac{\partial {f}_3}{\partial Q}
}
\]
where
\begin{align*}
f_1 &= \frac{k_3}{k_7} \left(k_1 + \frac{k_2K^2}{k_k^2 + K^2} \right) - \frac{k_{12}KQ}{\Gamma_k+K} - k_8K \\
f_2 &= \frac{k_5k_6/k_9}{1+(K/k_s)^5} - k_{10}S - k_{13}\Gamma_k\frac{SQ}{\Gamma_k+K} \\
	& \quad + k_{-13}(M_T-Q)\\
f_3 &= (k_{14}+k_{-13})(M_T-Q) - k_{13}\Gamma_k\frac{SQ}{\Gamma_k+K}
\end{align*}    
We can express $\delta \mathbf{X}(t)$ using the eigenvectors of $\mathbf{J}$ as a basis: $\delta \mathbf{X}(t) = \sum c_i{\bm\nu}_i e^{\lambda_it}$, where ${\bm \nu}_i$ are the right eigenvectors corresponding to eigenvalues $\lambda_i$, and $c_i$ are the components of $\mathbf{X}(t)$ along ${\bm \nu}_i$. Negative eigenvalues with large absolute values imply that deviations decay rapidly, and a widely separated set of eigenvalues enables us to eliminate these fast decaying modes. Fig.~\ref{fig:eigenphase} shows a plot of eigenvalues computed along the average trajectory for the 3D system described in (\ref{eq:threedimeq}) on a 10-base logarithm scaled polar coordinates.
\par
In order to compute the average trajectory, we sample the data from the Gillespie simulation of the full system; we then divide the trajectories into bins defined by their angular position with respect to an origin chosen in the centre of the trajectories (Fig.~\ref{fig:average_trajectory}). Next, we compute the mean eigenvalues for the Jacobian matrix computed in each bin as shown in Fig.~\ref{fig:polar_axis2}. In this figure, a particular eigenvalue is plotted in such a way that the distance from it to the origin is computed by taking a 10-base logarithm of its inverse absolute value. 
\par
It is clear that the 3 eigenvalues are separated from each other during the excitable state back to the vegetative state, making possible a reduction to a lower-dimensional system. In our case, the most negative eigenvalues are about 10 times as large as the others in absolute value, implying the existence of a low-dimensional attracting manifold. However, there also exists positive eigenvalues marked in Fig.~\ref{fig:eigenphase} which is the hallmark of an excitable system. For this reason, the whole space is divided into regions which are defined by positive and negative eigenvalues; the regions where the positive eigenvalues are found are demarcated by angles $\alpha$, $\beta$. In these regions, the trajectories could diverge along the direction of the eigenvectors, making the time series of the reduced and high-dimensional model different from each other, while the phase plane trajectories get pulled back to coincide on the low-dimensional manifold. 
\par
Since we are interested in the dynamical behaviour of the reduced model on $(K,S)$ space, we assume that $Q$ can be described as a function of $K$ and $S$, that is we assume $Q = Q(K,S)$. This means the dynamics of the system always lies close to a 2D manifold in $(K,S,Q)$ space and its velocity is uniquely determined by $K$ and $S$ alone. As a result, we have $\frac{dQ}{dt} = \frac{\partial Q}{\partial K}\frac{dK}{dt} + \frac{\partial Q}{\partial S}\frac{dS}{dt}$. Plugging this back into (\ref{eq:threedimeq}) we obtain $Q = F(K,S,\frac{\partial Q}{\partial K},\frac{\partial Q}{\partial S})  $ (The functional form of $F$ can be found in Appendix \ref{subsec:finding_q}).
\par
\begin{figure}[h]
\centering
		\includegraphics[width=0.5\textwidth]{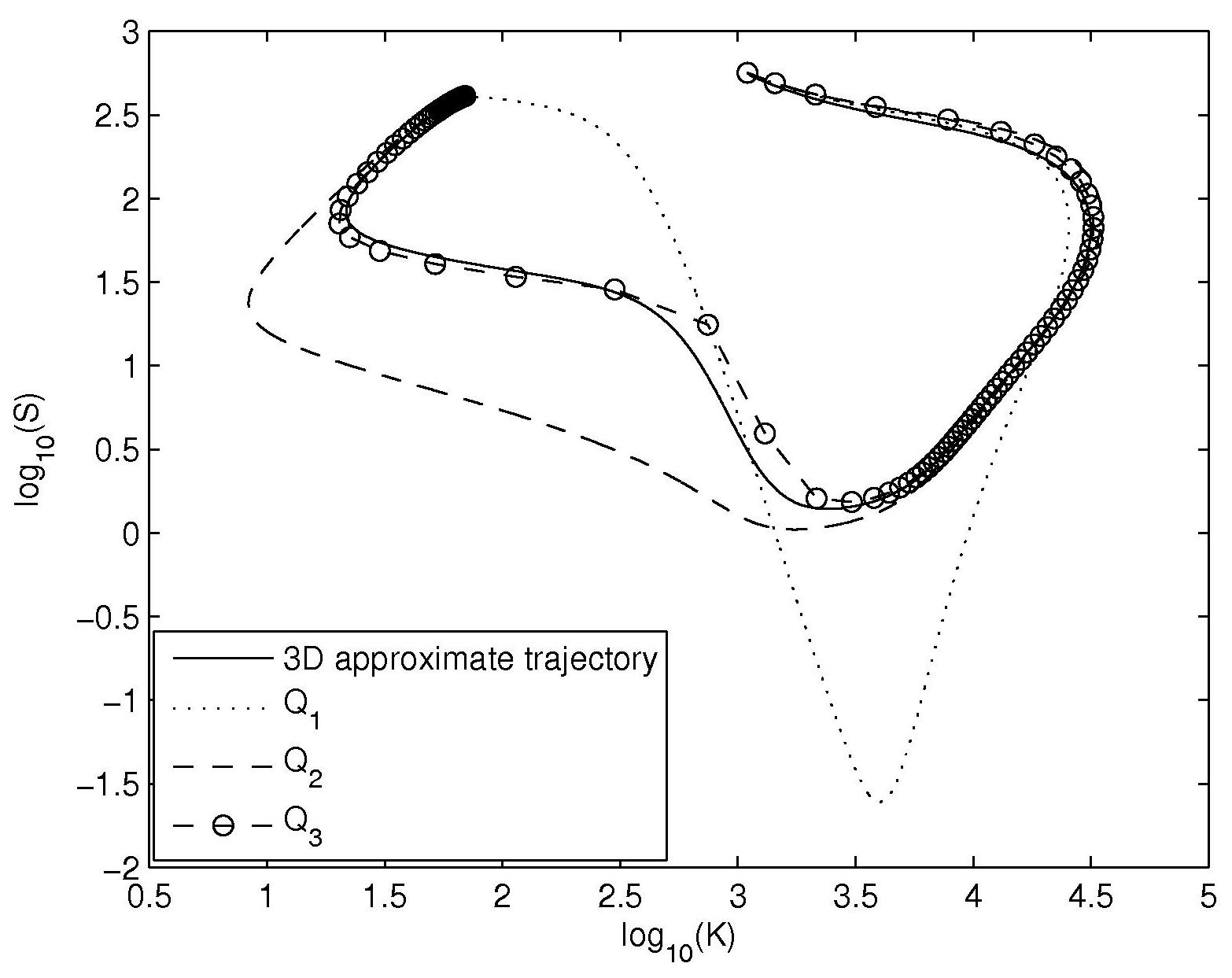}
	\caption{Trajectories generated by the 3D model and its 2D approximations ($Q_1$,$Q_2$,$Q_3$).}
	\label{fig:modelcomp}
\end{figure}

In order to estimate function $Q$, we use an iterative procedure~\citep{Fraser88,Roussel97,Roussel01} to define a sequence of approximations $Q_n$, starting with a trial function $Q_0(K,S) = 0$. Iteratively, we compute $Q_{n+1} = F(K,S,\frac{\partial Q_{n}}{\partial K},\frac{\partial Q_{n}}{\partial S})$, $n = 0,1,\dots$. Each step of the iteration $Q_n$ can be put back into (\ref{eq:threedimeq}) to obtain a 2D deterministic approximate model. Numerical experiments show that $Q_n$ converges rapidly and even $Q_2$ gives a very good approximation. In particular, Fig.~\ref{fig:modelcomp} shows the three different 2D models corresponding to the three first approximate functions of $Q$ ($Q_1$,$Q_2$,$Q_3$) compared with the 3D model. It turns out that $Q_1$ is the same expression as that obtained by setting $dQ/dt\approx 0$, the 2D adiabatic approximation. It is clear that the third approximation $Q_3$ very closely fits the 3D approximate model; therefore, we will take $Q_3$ as the deterministic approximation to the full system.
\begin{figure}[h]
\centering
	\includegraphics[width=0.5\textwidth]{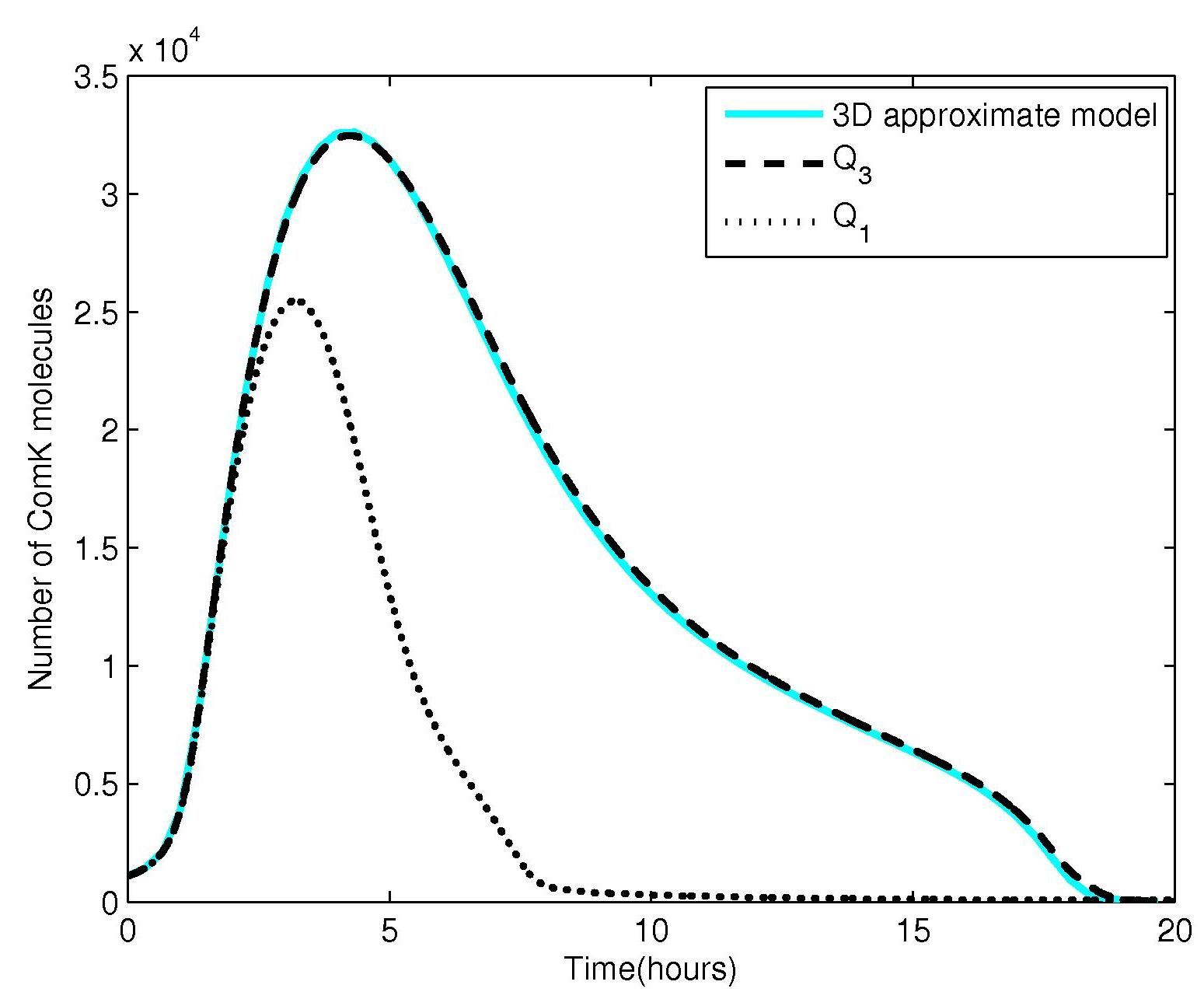} 
	\caption{Competence duration using $Q_3$ (the 2D approximation) and $Q_1$ (the adiabatic approximation) in comparison with the 3D approximate model. Note, the curves of $Q_3$ and the 3D approximation fall on top of each other.} 
	\label{fig:comp_duration_2DDeapprsys}
\end{figure}
\par
Fig.~\ref{fig:comp_duration_2DDeapprsys} shows a comparison of the competence durations between the naive adiabatic approximation introduced in~\citep{Suel06,Suel07} and the iteratively produced model where $Q_3(K,S)$ is used in the $K, S$ evolution equations. Evidently, the competence duration in the 2D approximation ($Q_3$) is about ten hours which agrees with that in the full system whereas this duration is only roughly four hours in the adiabatic approximation. This significant discrepancy implies that the naive adiabatic model provides a poor approximation of the original system. However, even though the 2D approximate model gives a much better approximation where the competence duration has been preserved, we still need to verify if this model gives similar stationary probability distribution by looking for a reduced stochastic model, which we address in the following section. 
\section{THE REDUCED STOCHASTIC MODEL}
\label{sec:redu_sto_model}
The CME was set up to capture two key properties that can be measured experimentally using fluorescent tagging. These are the time spent in the competence state and the probability of becoming competent. For our reduced 2D approximation to provide a good model of the full system, we want it to capture these same properties. To accomplish this, we want to make the probability density function (PDF) in the $K$-$S$ plane to be similar for both the 2D approximation and the CME. Having obtained a procedure for obtaining a 2D model for the time duration, we now move on to construct a set of stochastic differential equations (SDE), also called Langevin equations \citep{Gill02,Gill07}, that reproduce the histograms of the $K, S$ proteins obtained from the Gillespie simulation. For the duration of competence events, we assumed that our system --- the full model and the approximation --- was initialized to start in the competent regime. The introduction of stochasticity is designed to drive the transition from the stationary vegetative state to the competent state via fluctuations in species numbers, since the process of transcription and translation are stochastic events~\cite{Elowitz02}.
\par
\begin{figure}[h]
\centering
	\includegraphics[width=0.5\textwidth, height = 5cm]{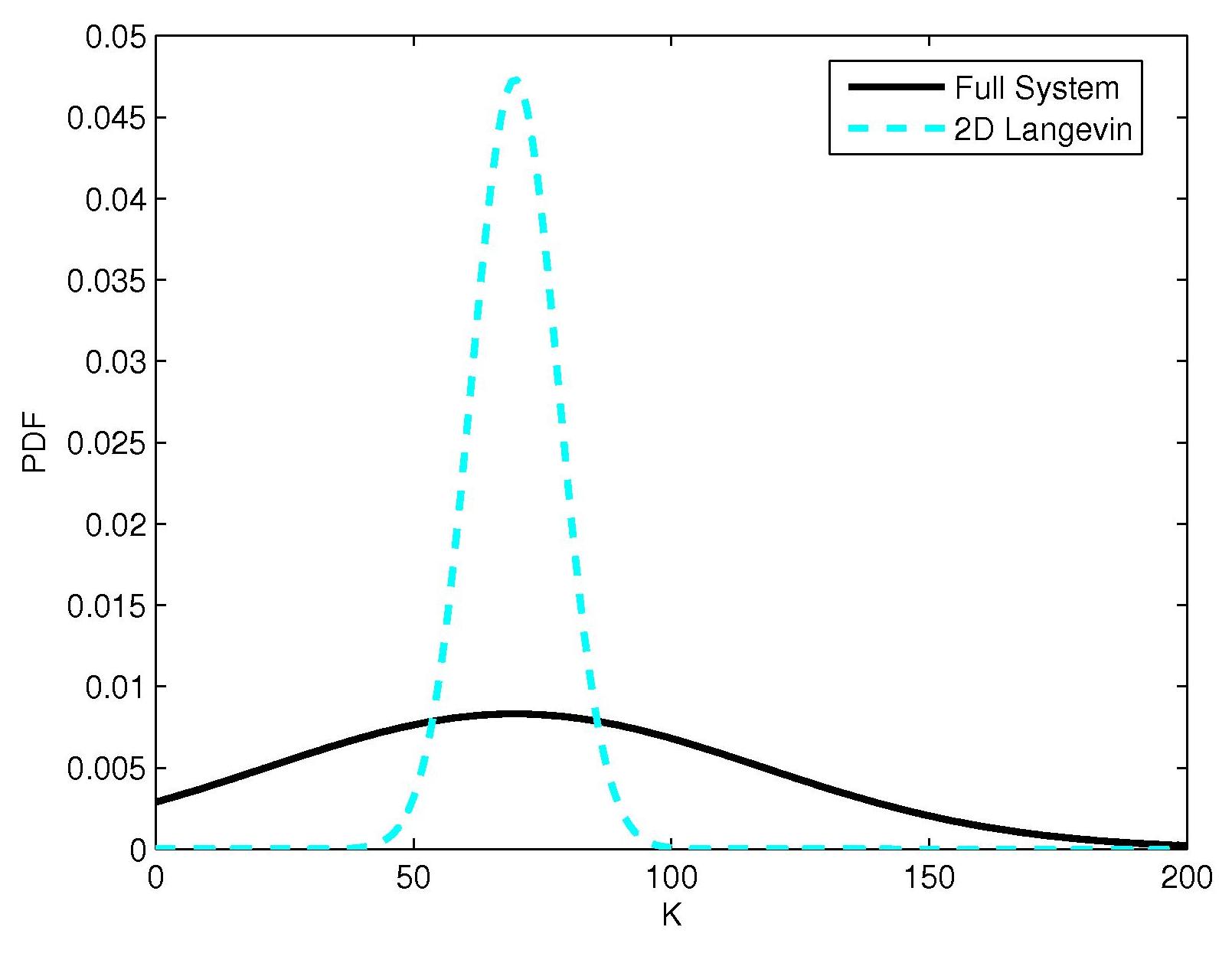}
	\caption{Comparison between the full system and the 2D Langevin model in terms of probability density function (PDF) of ComK. In both models, we sample the simulation data for $K$ and $S$ satisfying $0 \leq K \leq 200$ and $0 \leq S \leq 1000$. The PDF of ComK for each model is then computed and compared.} 
	\label{fig:2d_7d_pdf}
\end{figure}

The standard way of reducing a CME to a Langevin description can be described in two steps~\cite{Gill07}. Firstly, we replace the Poisson distributions implicit in the CME description with Gaussian distributions for the multiplicative fluctuations in the Langevin equation. Secondly, we adiabatically eliminate the fast reactions (as before) to obtain a reduced set of equations describing the dynamics of slowly varying species. For the competence regime, the elimination of the mRNA species did not introduce any errors in the dynamics. However, we find the variance in the noise is not linear in $K$; therefore, the standard assumption that the variance is proportional to the number of reactions is wrong. Indeed, Fig.~\ref{fig:2d_7d_pdf} shows the significant discrepancy in the dynamical behaviour between the 2D Langevin model and the full system near the steady state. The reason for this, we believe, is: the transition to the competence state is driven by small number of mRNA molecules subject to the $K^2$ non-linearity in the activating dynamics (see the $R_K$ evolution in (\ref{eq:7d_deterministic_model}), for $K\ll k_k$), rendering the replacement of the Poisson by a Gaussian distribution suspect. Furthermore, the tails of the distribution of $R_K$ molecules are drawn away, not back to the attracting steady state, via the repulsive dynamics that is characteristic of the excitable system (see the range of positive eigenvalues in Fig.~\ref{fig:polar_axis2}). In addition, even at the steady state fixed point the Jacobian is a non-normal matrix, and its non-orthogonal eigenvectors give rise to large transients driving transitions to the competent regime in this excitable system. Consequently, the elimination of the mRNA species of ComK leads to loss of the fluctuations needed to drive the competence behaviour. This is further reinforced by the observation that the linear noise (Gaussian) approximation around the steady state fails to account for the tails of the distribution around the steady state.  
\par
By tracking the noise-driven transition in a 2-species feedback circuit which exhibits the similar behaviour around the steady state, we found that the variance of the noise term is proportional to the square of the molecular number (see Appendix \ref{subsec:bistablemodel}). This finding therefore motivates us to introduce a Langevin model with a multiplicative noise extension of the 2D approximate ODEs from the previous section. This 2D stochastic model can be described as follows,
\begin{align}
\label{eq:2dstochastic}
\begin{split}
dK &= f_k(K,S,Q_3(K,S)) dt + \mu_k dw_k\\
dS &= f_s(K,S,Q_3(K,S)) dt + \mu_s dw_s
\end{split}
\end{align}      
where the functional forms $f_k$ and $f_s$ are given in (\ref{eq:threedimeq}), we introduce Wiener processes $dw_k$ and $dw_s$, and set $\mu_k = \sigma_k K$, $\mu_s = \sigma_s S$. This is because we are only interested in the tail of the stationary probability distribution where the noise terms are supposed to be proportional to $K$ and $S$. The magnitudes $\sigma_k$, $\sigma_s$ of the noise terms are chosen to obtain the dynamical behaviour qualitatively similar to that of the CME. Additionally, the initiation probability of competent events computed from the stochastic model should also quantitatively be preserved. However, this probability is very sensitive to the switching behaviour driven by the noise terms. In other words, a slight change in coefficients $\sigma_k$, $\sigma_s$ leads to a significant change in the initiation probability. This is because of the exponential sensitivity of the tail of the probability distribution to the noise-driven switching state in our particular circuit~\cite{Mehta08}. Consequently, we have tried the simulation with different values of $\sigma_k$, $\sigma_s$ and found that the stochastic model which provide a relatively good approximation has $\sigma_k = 0.008$, $\sigma _s = 0.005$ (see Appendix \ref{subsec:optimization}). We show in Fig.~\ref{fig:reduced_trajectories} trajectories generated by the Euler-Maruyama method \cite{Higham01} from the 2D stochastic model described by (\ref{eq:2dstochastic}), which are similar to those generated from the CME by the Gillespie algorithm (Fig.~\ref{fig:trajdismod}). In order to obtain the stationary probability distribution, we solve the Fokker-Planck equation \cite{Gill02} which is described as follows:
\begin{align}
\label{eq:fokker_planck}
\frac{\partial P(K,S,t)}{\partial t} &= -\frac{\partial}{\partial K}f_k(K,S,Q_3(K,S))P(K,S,t) \\ \nonumber
& \quad - \frac{\partial}{\partial S}f_s(K,S,Q_3(K,S))P(K,S,t)   \\ \nonumber
& \quad + \frac{1}{2}\frac{\partial^2}{\partial K^2}\mu_k^2P(K,S,t) + \frac{1}{2}\frac{\partial^2}{\partial S^2}\mu_s^2P(K,S,t)
\end{align}
\begin{figure}[h!]
\centering
	\includegraphics[width=0.5\textwidth]{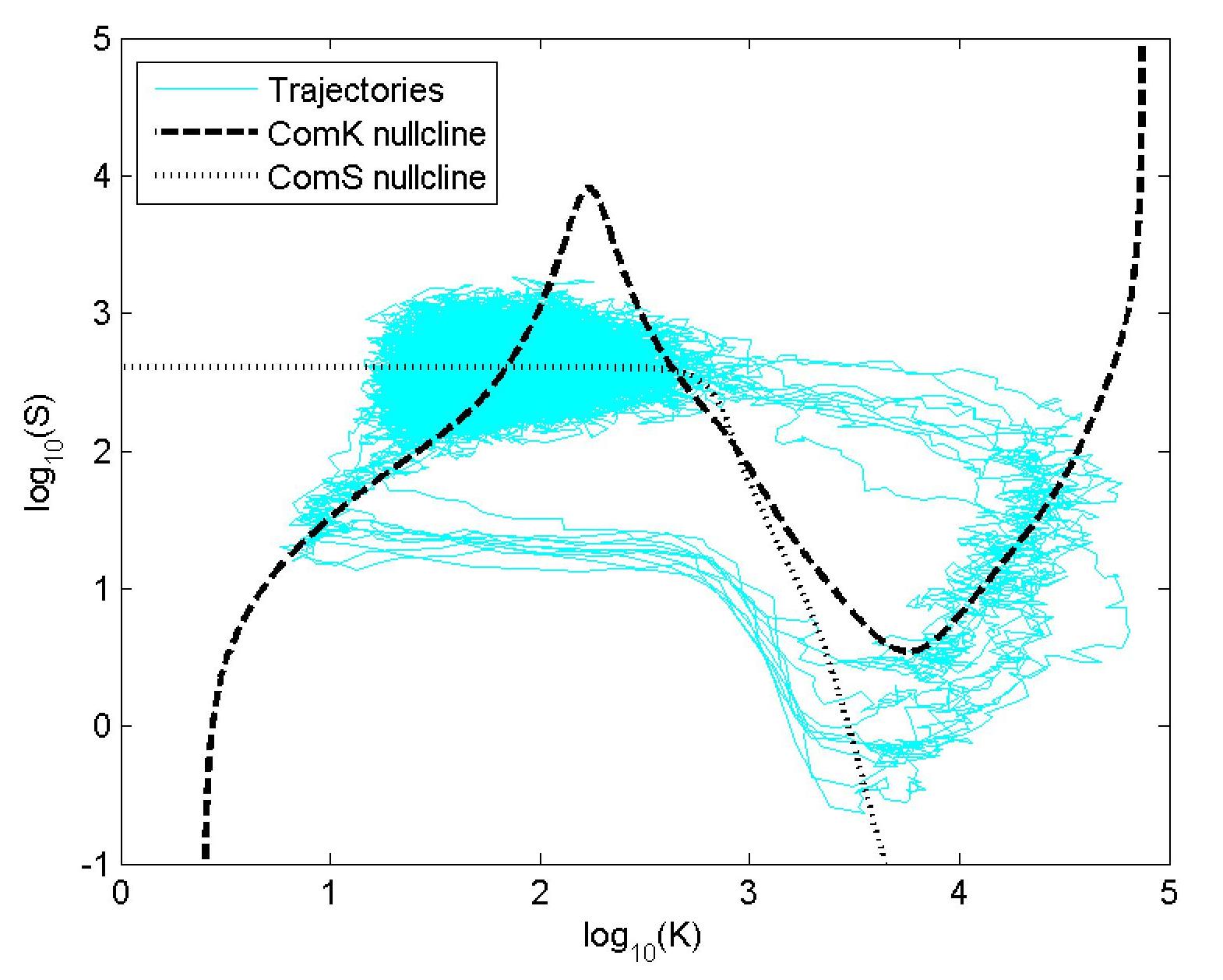}
	\caption{Trajectories generated by the 2D stochastic model given by (\ref{eq:2dstochastic}) with coefficients $\sigma_k = 0.008$, $\sigma_s = 0.005$. The nullclines of the 2D approximate model ($Q_3$) are also plotted.}
	\label{fig:reduced_trajectories}
\end{figure}
\par
\begin{figure*}
	\centering
	\mbox{
	\subfigure[][]
	{
		\includegraphics[width=0.5\textwidth]{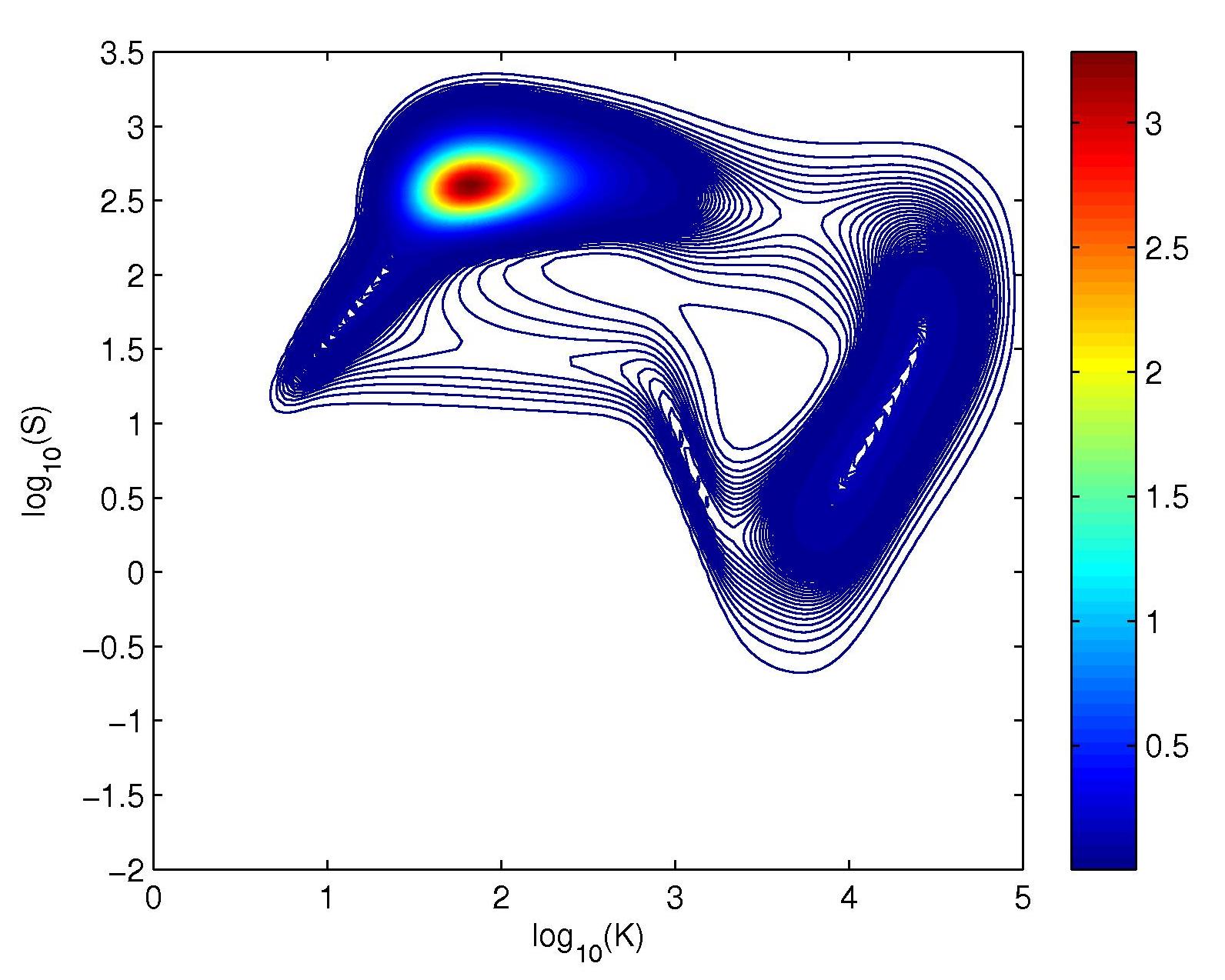}
		\label{fig:approxfokplnc}
	} \quad
	\subfigure[][]
	{
			\includegraphics[width=0.5\textwidth]{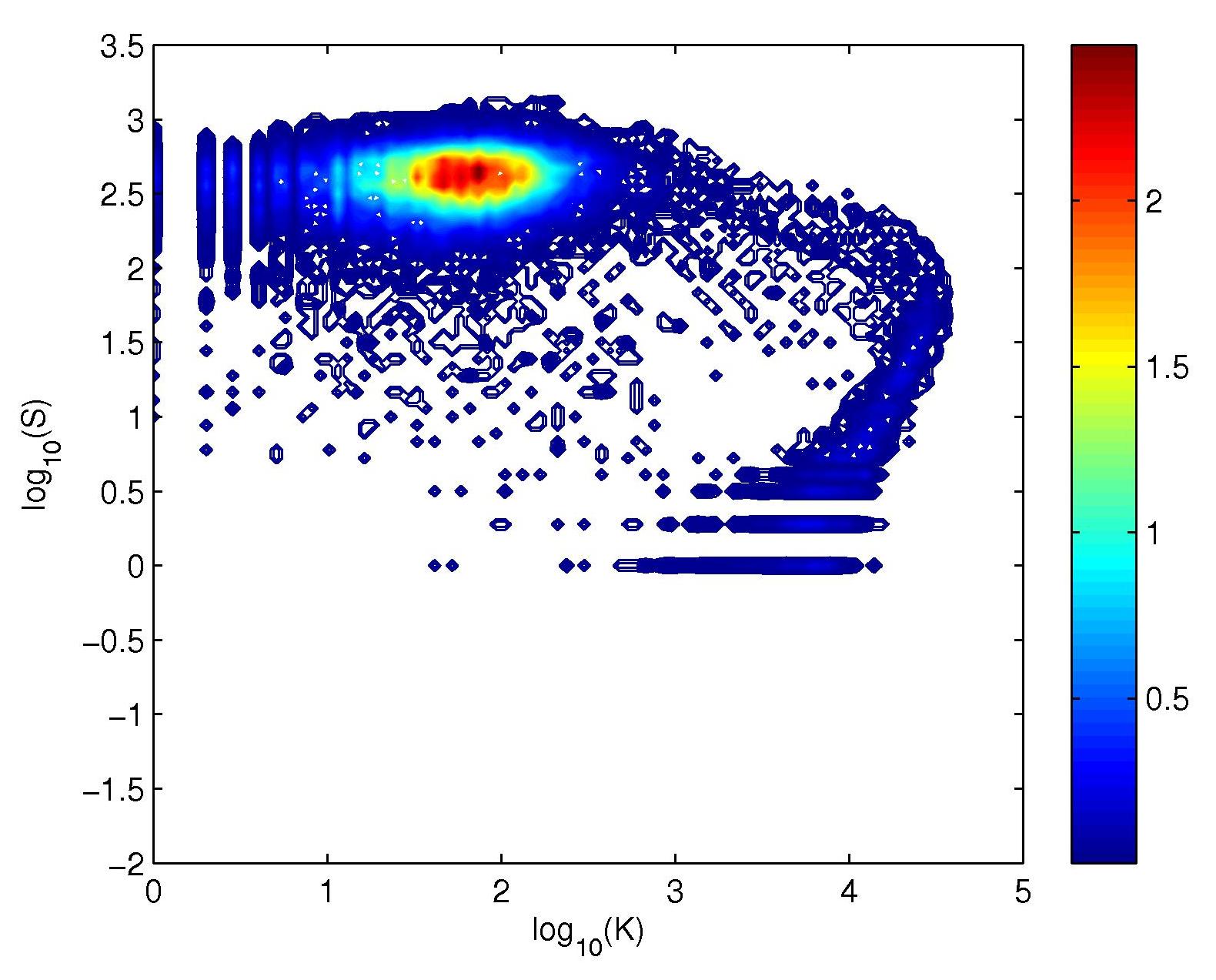}
		 \label{fig:3d_wild_type_probability_distribution}
	}}
	\caption{Probability density functions (PDFs) of the phenomenological stochastic 2D model \subref{fig:approxfokplnc} and the full discrete model \subref{fig:3d_wild_type_probability_distribution}.}
\label{fig:probdis_comparison}
\end{figure*}

Fig.~\ref{fig:probdis_comparison} shows the probability density function (PDF) of the 2D reduced model in comparison with that computed from the full model. As we can see, both models produce similar bimodal distributions that are characteristic of the cell counts in the vegetative and competence states obtained in~\cite{Suel06}. 
\section{CONCLUSION}
In this paper we have systematically conducted a series of customary approximations of the chemical master equations of the excitable model
of competence and found many to be not faithful to the dynamics of the
full model.  First, to obtain the ODE description of \cite{Suel07} we
replaced the stochastically varying quantities by their mean values and
further assumed that the averages of nonlinear functions $\langle
f(x)\rangle$ of a stochastic variable $x$ can be replaced by non-linear
functions of the averaged variable $f(\langle x \rangle)$. The
reduction of the ODEs thus obtained to a two-dimensional set is usually
accomplished by dividing the dynamical processes into a set of fast and
a set of slow reactions. If the variables are also divided in a similar
way into a fast and slow set, a fast-slow decomposition of the dynamics
is easily facilitated. This allows the slow variables to be viewed as
unchanging when considering the fast dynamics; the slow variables are
only affected by time averages of the fast variables. This adiabatic
approximation --- replacing fast variables by their equilibrium values ---
turns out to be useful in retaining the same fixed points as the high
dimensional deterministic dynamical system, and was used as the basis of
the stability analysis in \cite{Suel07}. We find, however, that the fast
reactions do not segregate the fast and slow species in the model
description. We propose a suitable functional dependence between the
variables that allows the system to be described by a slow set of
variables that adequately matches the time scales of the competence
regime, unlike the standard adiabatic approximation of \cite{Suel07}.
\par
In accounting for the noise-driven transition to the competence regime,
we needed to put back the noise that the mRNA species contributes to
this process. At the vegetative steady state, the balance between the
protein production terms and the linear decay term might suggest that
the variance of the Poisson distribution be proportional to the protein
concentration. By explicit simulation of the ComK mRNA-protein subset,
we find this does not hold, and instead find that the standard deviation
is linear in the protein concentration. Using this as a guide, we
introduce a 2-dimensional stochastic differential equation with a Wiener
process proportional to a linear term in the concentration, and fit the
constant of proportionality by minimising the distance between the
distributions obtained from the chemical master equation and the
2-dimensional approximation. This approximate 2-dimensional system
could then form the basis of future studies on the dependence of first
passage time distributions on the system parameters, and other
theoretical characterisations. We propose that this indirect way of
approximating the behaviour of a high dimensional chemical system by a
low-dimensional set of stochastic differential equations could be used
as a generic method when direct ways of model reduction fail.
\appendix
\subsection{Derivation of Eqs. (\ref{eq:threedimeq})}
The first model reduction step would be eliminating both mRNAs in Eqs. \eqref{eq:7d_deterministic_model} by setting $dR_{K,S}/dt = 0$, we obtain
\begin{align}
R_K &= \frac{k_1 + \frac{k_2K^2}{{k_k}^2 + K^2}}{k_7}\label{eq:r_k} \\ 
R_S &= \frac{k_5}{1 + (K/k_s)^5}{k_9}\label{eq:r_s}
\end{align}
On the other hand, we also have,
\begin{align}
M_K &= \frac{K}{\Gamma_k + K}Q \label{eq:m_k} \\ 
M_S &= M_T - M_K - A = M_T - Q \label{eq:m_s} \\ 
A &= Q - M_K = Q - \frac{K}{\Gamma_k + K}Q = \frac{\Gamma_k}{\Gamma_k + K}Q \label{eq:a}
\end{align}
From Eqs. (\ref{eq:r_k}), (\ref{eq:m_k}) and (\ref{eq:a}), we have
\begin{align}
\label{eq:dkdt}
\frac{dK}{dt} &= -k_{11}KA + k_{-11}M_K + k_3R_K - k_8K \\ \nonumber
&= -k_{11} \Gamma_k \frac{KQ}{\Gamma_k + K} + k_{-11} \frac{KQ}{\Gamma_k + K}  \\ \nonumber
& \quad + \frac{k_3}{k_7} \left(k_1 + \frac{k_2K^2}{k_k^2 + K^2} \right) - k_8K \\ \nonumber
&= \frac{k_3}{k_7} \left(k_1 + \frac{k_2K^2}{k_k^2 + K^2} \right) - \frac{k_{12}KQ}{\Gamma_k+K} - k_8K
\end{align}
where $k_{12} = k_{-11} - k_{11}\Gamma_k$.
Similarly, from (\ref{eq:r_s}), (\ref{eq:m_s}) and (\ref{eq:a}), we have
\begin{align}
\label{eq:dsdt}
\frac{dS}{dt} &= -k_{13}SA + k_{-13}M_S + k_6R_S - k_{10}S \\ \nonumber
&= \frac{k_5k_6/k_9}{1+(K/k_s)^5} - k_{10}S - k_{13}\Gamma_k\frac{SQ}{\Gamma_k+K} \\
	& \quad + k_{-13}(M_T-Q)\\
\end{align}
Finally,
\begin{align}
\label{eq:dqdt}
\frac{dQ}{dt} &= -\frac{dM_S}{dt} \\ \nonumber
&= (k_{-13}+k_{14})M_S - k_{13}SA \\ \nonumber
&= (k_{-13}+k_{14})(M_T - Q) - k_{13}\Gamma_k\frac{SQ}{\Gamma_k+K} 
\end{align}
From (\ref{eq:dkdt}), (\ref{eq:dsdt}) and (\ref{eq:dqdt}), we obtain (\ref{eq:threedimeq}).
\subsection{Iterative Procedure For Finding $Q$}
\label{subsec:finding_q}
In order to find the explicit form of $Q$, we rewrite the equation as follows:
\begin{align*}
&(k_{14}+k_{-13})(M_T-Q) - k_{13}\Gamma_k\frac{SQ}{\Gamma_k+K} \\
&= \frac{dQ}{dK}\left(\frac{k_3}{k_7} \left(k_1 + \frac{k_2K^2}{k_k^2 + K^2} \right) - \frac{k_{12}KQ}{\Gamma_k+K} - k_8K\right) \\
& \quad + \frac{dQ}{dS}\Bigg(\frac{k_5k_6/k_9}{1+(K/k_s)^5} - k_{10}S - k_{13}\Gamma_k\frac{SQ}{\Gamma_k+K} \\
& \quad + k_{-13}(M_T-Q)\Bigg)
\end{align*}
hence,
\begin{equation*}
Q = F\left(K,S,\frac{dQ}{dK},\frac{dQ}{dS}\right) = \frac{A-B}{C-D}
\end{equation*}
where
\begin{align*}
A &= (k_{14}+k_{-13})M_T - \frac{dQ}{dK}\left(\frac{k_3}{k_7}\left(k_1 + \frac{k_2K^2}{k_k^2+K^2}\right) - k_8K\right) \\
B &= \frac{dQ}{dS}\left(k_{-13}M_T + \frac{k_5k_6/k_9}{1+(K/k_s)^5} - k_{10}S\right) \\
C &= \frac{k_{13}\Gamma_k S}{\Gamma_k + K} + k_{14} + k_{-13} - \frac{dQ}{dK}\frac{k_{12}K}{\Gamma_k + K} \\
D &= \frac{dQ}{dS}\left(\frac{k_{13}\Gamma_k S}{\Gamma_k + K} + k_{-13}\right)
\end{align*}
\subsection{Reduction of a 2-Species Model to Track Noise-driven Transition}
\label{subsec:bistablemodel}
As we mentioned early in section \ref{sec:redu_sto_model}, the usual Poissonian noise in the Langevin approximation turned out to be incorrect. In order to address the source of this issue, we will focus our attention on the behaviour of the system near the stable fixed point and the transition beyond the intermediate unstable fixed point to the competent state. To do this, we will be looking at a much simpler noise-driven switching circuit which is extracted from the original system. This is done by just looking at the dynamics of two variables ComK and mRNA while ignoring the effect of the other variables, this will give us a bistable model. Even though the behaviour of this bistable model is different from the full model in a long period of time, the dynamics of the system near the fixed point should be similar in both models. Hence, if we can capture the noise near the transition which is believed to account for the tails of the distribution around the steady state, we then can construct a stochastic model which describes the similar behaviour as that in the full system.
\par
The two-species model is obtained by only considering the chemical reactions where a protein activates its own transcription as follows:
\begin{center}
	\begin{align}
	\label{eq:simplemodel}
	\begin{split}
	\cee{
	{\rm P_{comK}^{const}} &\xrightarrow{k_1} {\rm P_{comK}^{const}} + { \rm mRNA_{comK}}  \\     
	{\rm P_{comK}} &\xrightarrow{f\left( [ComK],k_2,k_k\right)} {\rm P_{comK}} + {\rm mRNA_{comK}} \\
	{\rm mRNA_{comK}} &\xrightarrow{k_3} {\rm mRNA_{comK}} + {\rm ComK} \\
	{\rm mRNA_{comK}} &\xrightarrow{k_4} \emptyset \\
	{\rm ComK} &\xrightarrow{k_6} \emptyset 
	}
	\end{split}
  \end{align}
\end{center}
where $f\left( [ComK],k_2,k_k\right) = \frac{k_2[ComK]^2}{{k_k}^2 + [ComK]^2}$.
\par
The first two reactions represent how much $mRNA_{comK}$ is produced from the binding of protein to the promoters on DNA. The next reaction shows how much protein ComK is synthesized from $mRNA_{comK}$. The fourth and fifth reactions represent the linear degradation of the mRNA and protein, respectively.  
In fact, this model is simplified from the 7D model by setting the variables $MecA$ and $MecA_K$ to their steady values. Moreover, the system exhibits bistability and transition from a low to a high expression state of ComK which is driven by noise in mRNA levels.
\par
We denote the protein and mRNA as $K$ and $m$, respectively. As a result, the deterministic differential equations for this model are described as follows:
\begin{equation}
\begin{split}
\label{eq:deter2dmodel}
\frac{dK}{dt} &= k_5 + k_3m - k_6K \\
\frac{dm}{dt} &= k_1 + \frac{k_2K^2}{{k_k}^2+K^2} - k_4m
\end{split}
\end{equation}
The model parameters are given in Table \ref{tbl:simple_model_parameters}, here we introduce the parameter $k_5$ ($k_5 = 3.24 \times 10^{-5}$) in order to keep the structure and the position of the fixed points the same as that in the original 7D system.
\begin{table}
\caption{
	Model parameters
	}
 \label{tbl:simple_model_parameters}
\begin{center}
   \begin{tabular}{|c||c|} \hline
   $k_1$& $0.00021875 s^{-1} \, nM$    \cr \hline
   $k_2$ & $0.1875 s^{-1}$ \cr \hline
   $k_3$ &   $0.2 s^{-1}$ \cr \hline
   $k_4$ & $0.005 s^{-1}$  \cr \hline
	$k_5$ & $3.2 \times 10^{-5} s^{-1}$  \cr \hline
	$k_6$ & $1.4704 \times 10^{-4} s^{-1}$  \cr \hline
	$k_k$ & $5000 \, nM$  \cr \hline
   \end{tabular}
  
\end{center}
\end{table}
\par
In this model, we found that the 2D Langevin simulation breaks down due to the very small number of mRNA population having been driven to negative values. On the other hand, we notice that the mRNA lifetimes are shorter than protein lifetimes ($\frac{k_4}{k_6} = 34 \gg 1$), this therefore motivates us to adiabatically eliminate this small variable to obtain a 1D Langevin model. However, the simulation result shows that this model remains around the steady state and never goes to the high expression regime. This means the adiabatic approximation does not capture the noise-driven transitions in this model, and the fluctuation in the mRNA which is generated in the 2D Langevin model significantly contributes to the switching behaviour of the system. Moreover, the mRNA reduction is also problematic due to the fact that the time scales of mRNA and protein are not completely separated (despite the decay rate of mRNA being 34 times faster than that of protein, the production rate of mRNA ($k_3 = 0.2$) is about 1000 times faster than the decay rate of protein ($k_6 = 1.4704 \times 10^{-4}$)). Even though the mRNA reduction issue can prevent us from coming up with a good approximate model, there is a way around it. In particular, we will estimate the size of fluctuation near the steady state in the 2D Gillespie model whereby we would hope to construct the correct noise for the stochastic model.
\begin{figure}
\centering
		\includegraphics[width=0.5\textwidth]{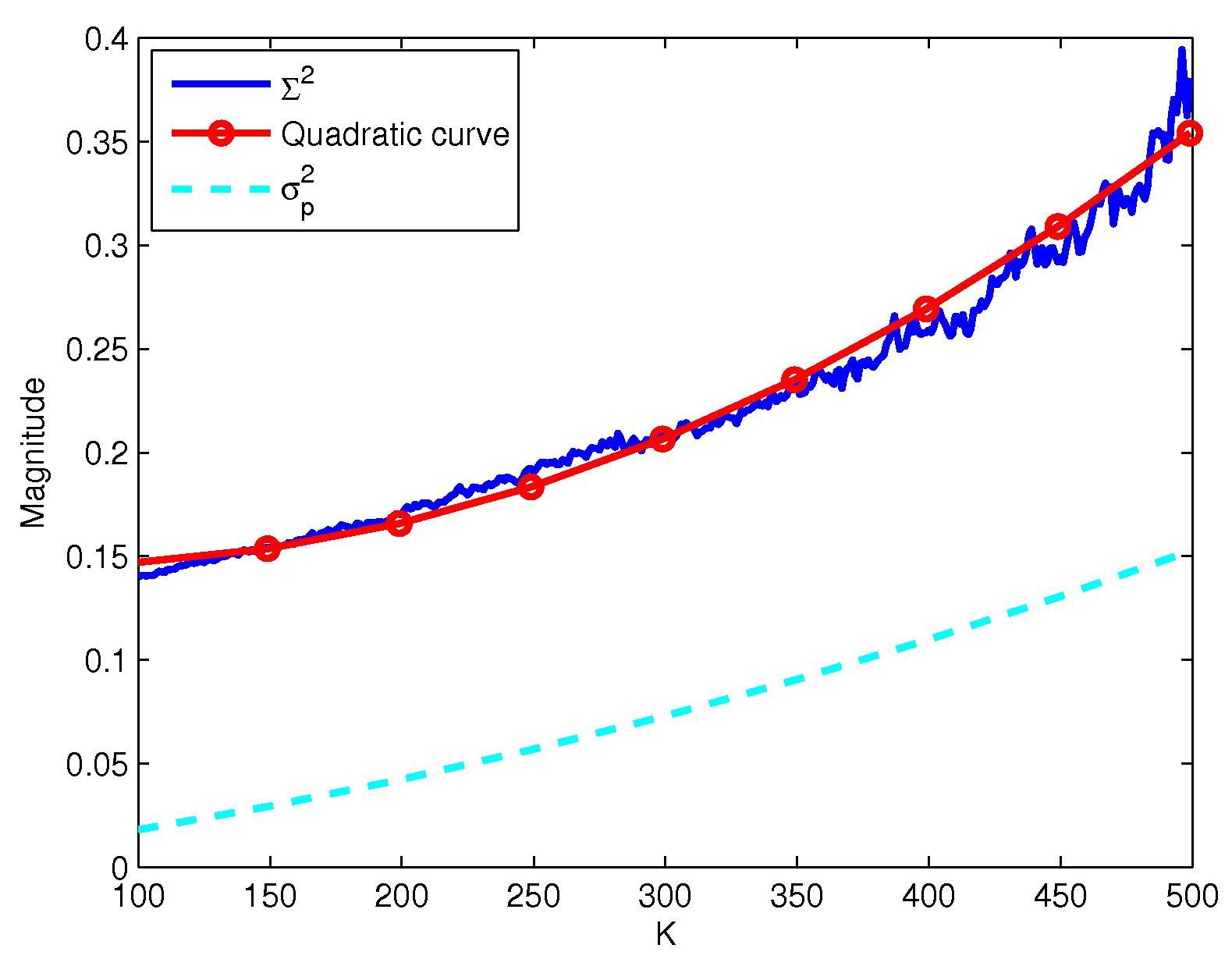}
	\caption{The square of size of fluctuation ($\Sigma^2$) is well fitted by a quadratic curve. This means the noise term in the Langevin model should be proportional to the number of ComK. The discrepancy between the variance of Poisson noise ($\sigma^2_p$) and the empirical noise ($\Sigma^2$) is also shown.}
	\label{fig:simple_model_fluctuation_size}
\end{figure}
\par
In order to estimate the size of fluctuation, we run Gillespie simulations using the reaction scheme given by \ref{eq:simplemodel} with the initial condition given by the fixed point. For each run, we stop the simulation as soon as the molecular number of protein exceeds 500. This is the threshold over which the system enters the high expression state. Next, we collect and put the simulation data into $500$ separate bins according to different values of protein $K$ (notice that we are only interested in the values of the protein and mRNA, not the time step). In particular, each bin $i = 1,2, \dots 500$ contains a particular value of protein $K_i$ and a set of all possible values of $mRNA$ with respect to $K_i$ ( we don't count the frequency of $mRNA$). Let $L_i$ be the total number of $mRNA$ values in bin $i$, then the value of an instance of mRNA $j$ belonging to bin $i$ is denoted as $m_{ij}$ where $j = 1, 2, \dots L_i$. For each bin $i = 1,2, \dots 500$, we compute the expected change in protein $K_i$ in time step $\Delta t$ denoted as $\Delta K_i$ that is determined by the propensity functions in which the protein gets involved. According to this, the expected change in $K_i$ given $m_{ij}$ in a time interval $\Delta t$ denoted as $\Delta K_i^j$ can be estimated as $\Delta K_i^j = (k_5 + k_3m_{ij} - k_6K_i) \Delta t$. In our case, we take $\Delta t$ to be the same as that in the 1D Langevin model. Since $\Delta t$ is the same in both Gillespie and 1D Langevin models, the only comparable term would be $k_5 + k_3m_{ij} - k_6K_i$. Thus, we can ignore $\Delta t$ and re-define $\Delta K_i^j$ as follows:
\begin{equation}
\Delta K_i^j \equiv k_5 + k_3m_{ij} - k_6K_i
\end{equation}  
where $j = 1,2, \dots L_i$. The variance of $\Delta K_i$ is then computed by the following equation:
\begin{equation}
\sigma_{\Delta K_i}^2 = \frac{1}{L_i}\sum_{j=1}^{L_i}(\Delta K_i^j - \left\langle\Delta K_i\right\rangle)^2 
\end{equation}
Here, $\left\langle\Delta K_i\right\rangle = \frac{1}{L_i}\sum_{j=1}^{L_i} \Delta K_i^j$. On the other hand, since CME can be described by a Poisson process; therefore, the variance in protein caused by this process is given as follows:
\begin{equation}
\Sigma_i^2 = k_5 + k_3\left<m_i\right> + k_6K_i 
\end{equation}
Here, the mean of mRNA for each bin $i$ is computed as
\begin{equation}
\label{eq:mean_mrna}
 \left\langle m_i\right\rangle = \frac{1}{L_i}\sum_{j=1}^{L_i}m_{ij}
\end{equation}
As a result, the size of fluctuation for this particular data bin is given by 
\begin{equation}
\label{eq:fluctuation_size}
\Sigma = \sqrt{\Sigma_i^2 + \sigma_{\Delta K_i}^2} 
\end{equation}
By estimating the size of fluctuation in ComK ($\Sigma$), we found that this quantity is approximately proportional to $K$. Indeed, Fig.~\ref{fig:simple_model_fluctuation_size} shows the estimated square of size of fluctuation ($\Sigma^2$) in ComK which we can fit by a polynomial fitting curve. The fitting is done by minimizing the sum of the squares of the deviations of the data from the empirical curve (least-square fit). Since we are only interested in fitting the part of the curve which account for the tail of the probability distribution, we will do the fitting for $100 \leq K \leq 500$. In Fig.~\ref{fig:simple_model_fluctuation_size}, the empirical curve is well fitted by a quadratic curve which is defined as follows: 
\begin{equation*}
y_0 = b_2K^2 + b_1K + b_0
\end{equation*}
where $b_2 = 1.1 \times 10^{-6}$, $b_1 = -0.00014$, $b_0 = 0.15$. Also in Fig.~\ref{fig:simple_model_fluctuation_size}, the variance of Poisson noise ($\sigma^2_p$) when adiabatically eliminating the mRNA by setting $m = \frac{k_1 + \frac{k_2K^2}{{k_k}^2+K^2}}{k_4}$ is shown for comparison. This quantity is estimated as follows:
\begin{equation*}
\sigma^2_p = k_1 + k_3\left(\frac{k_1 + \frac{k_2K^2}{{k_k}^2+K^2}}{k_4}\right) + k_6K
\end{equation*}
As we can see in Fig.~\ref{fig:simple_model_fluctuation_size}, there is a significant difference between the variance of Poisson noise and the empirical noise ($\Sigma^2$). This implies that the Possion noise does not capture the correct noise in the Gillespie model which can only be fitted by a much bigger noise, and the magnitude of this noise is demonstrated by the quadratic fitting curve. We also notice that the empirical curve does not grow linearly to $K$; therefore, it should not be approximated by a linear line. This means the variance is proportional to the square of the molecular number ComK.
\subsection{Parameter Optimization Procedure}
\label{subsec:optimization}
\begin{figure}
\centering
	\includegraphics[width=0.5\textwidth]{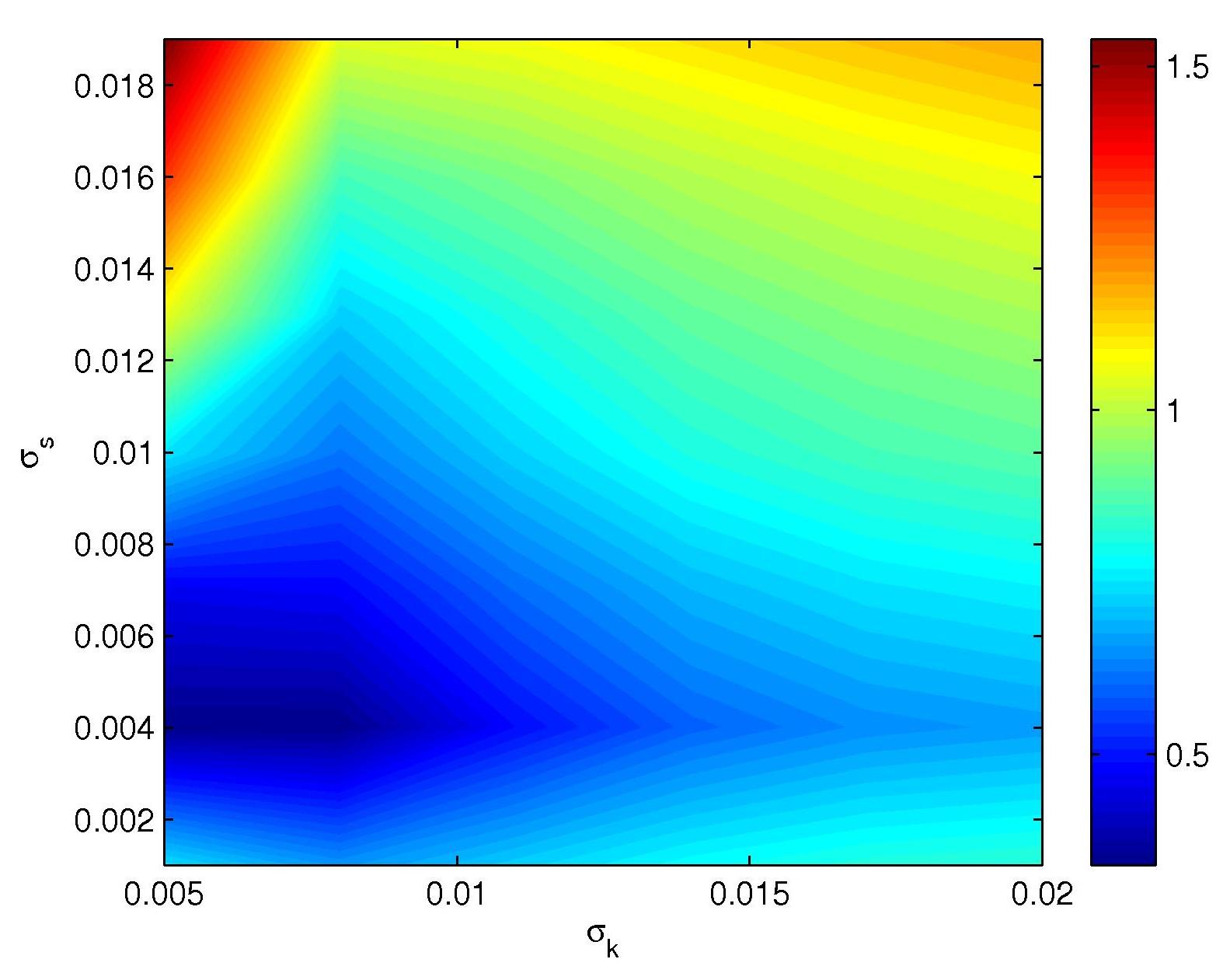}
	\caption{Distance as a function of $\sigma_k$ and $\sigma_s$.}
	\label{fig:KL_divergence_distance}
\end{figure}

As mentioned earlier, we will try to optimize the parameters $\sigma_k$, $\sigma_s$ such that the stationary probability distribution obtained from solving the Fokker-Planck equation (\ref{eq:fokker_planck}) close to that computed from the full discrete model. To do so, we use the Jensen-Shannon divergence \cite{Fuglede04} which is a smoothed version of the Kullback-Leibler divergence \cite{Johnson01} to measure the distance between the two distributions, says $P$ and $Q$. 
For each pair of parameters $(\sigma_k,\sigma_s)$, we compute the Jensen-Shannon divergence $D(P,Q)$ and sort them in descending order, we then choose the pair of parameters corresponding to the smallest distance. The pair of parameters chosen for our experiment satisfies $0.005 \le \sigma_k \le 0.02$ and $0.001 \le \sigma_s \le 0.02$. Fig.~\ref{fig:KL_divergence_distance} shows the distance between the PDFs in the full model and stochastic model as a function of the noise terms. In our experiment, we found that the stochastic model provides the best approximation with $\sigma_k = 0.008$, $\sigma _s = 0.005$.

\bibliographystyle{IEEEtran} 
\bibliography{IEEEabrv,ref}
\end{document}